\documentclass[aps,twocolumn,groupedaddress,showpacs]{revtex4}

\usepackage{epsfig,amsbsy,bm}

\newcommand{\Eref}[1] {Eq.~(\ref{#1})}
\newcommand{\Fref}[1] {Fig. \ref{#1}}

\newcommand{\be}{\begin{equation}}
\newcommand{\ee}{\end{equation}}

\newcommand{\ra}{\rangle}
\newcommand{\la}{\langle}

\renewcommand{\bbox}{\boldsymbol }

\newcommand{\hs}{\hspace*}
\newcommand{\vs}{\vspace*}

\newcommand{\bt}{\begin{tabular}}
\newcommand{\et}{\end{tabular}}
\newcommand{\bp}{\begin{minipage}}
\newcommand{\ep}{\end{minipage}}

\newcommand{\br}{\begin{eqnarray*}}
\newcommand{\er}{\end{eqnarray*}}
\newcommand{\ba}{\begin{eqnarray}}
\newcommand{\ea}{\end{eqnarray}}
\renewcommand{\r}{{\bbox r}}
\newcommand{\q}{{\bbox q}}
\newcommand{\ds}{\displaystyle}

\renewcommand{\k}{{\bbox k}} 

\newcommand{\z}{{\bbox z}} 
\newcommand{\E}{{\cal E}} 
 
\newcommand{\M}{{\cal M}} 
\newcommand{\Y}{{\cal Y}}

\newcommand{\nn}{\nonumber}

\newcommand{\isum}%
{\mathop{\hbox{$\displaystyle\sum\kern-13.2pt\int\kern1.5pt$}}}

\begin{document}

\bibliographystyle{apsrev}

\title
{Timing analysis of two-electron photoemission}

\author{ A. S. Kheifets$^{1}$\footnote[1]{Corresponding author: A.Kheifets(at)anu.edu.au}
}
\author{I. A. Ivanov$^1$}
\author{Igor. Bray$^2$}

\affiliation
{$^1$Research School of Physical Sciences,
The Australian National University,
Canberra ACT 0200, Australia
\\
$^2$ARC Centre for Matter-Antimatter Studies, Curtin University, WA 6845
Perth, Australia }

\date{\today}

\begin{abstract}
We predict a significant delay of two-electron photoemission from the
helium atom after absorption of an attosecond XUV pulse.  We
establish this delay by solving the time dependent Schr\"odinger
equation and by subsequent tracing the field-free evolution of the
two-electron wave packet.  This delay can also be related to the
energy derivative of the phase of the complex double photoionization
(DPI) amplitude which we evaluate by the convergent close-coupling
method.  Our observations prompt future attosecond streaking
experiments on DPI of He which can elucidate various mechanisms of
this strongly correlated ionization process.
\end{abstract}


\maketitle

The attosecond streaking has made experimentally accessible the
characteristic timescale of electron motion in atoms
\cite{Baltuska2003,Kienberger2004}.  Recent applications of this
technique to atomic photoionization, both in the near-infrared (NIR)
\cite{P.Eckle12052008} and extreme ultraviolet (XUV) 
\cite{M.Schultze06252010} spectral energy
range, revealed a noticeable time delay between subjecting an atom to
a short laser pulse and subsequent emission of the photoelectron.
While in the NIR photon energy range such a delay can be related to
nonadiabatic tunneling \cite{PhysRevA.63.033404}, the XUV delay can
be, at least partially, attributed to the energy dependent phase of
the complex photoionization amplitude
\cite{PhysRevLett.105.073001,PhysRevLett.105.233002}. 
This observation is particularly important as it allows for a complete
characterization of the photoionization process in a so-called
complete photoionization experiment \cite{0953-4075-37-6-010}.

In the case of double photoionization (DPI), all the essential
information on the many-electron dynamics of this strongly correlated
ionization process is contained in a pair of symmetrized {\em gerade}
and {\em ungerade} amplitudes. The moduli of these amplitudes and
their relative phase can now be determined experimentally
\cite{0953-4075-36-16-102,0953-4075-38-6-003}. In this Letter, we 
demonstrate that an additional information on the individual phases of the DPI
amplitudes can be supplemented by an XUV time delay measurement. In
our demonstration, we consider the helium atom driven by an XUV
attosecond pulse with the same parameters as employed in the
attosecond streaking experiment on Ne by \citet{M.Schultze06252010}.
By solving the time dependent Schr\"odinger equation (TDSE) and by
tracing subsequent field-free evolution of the two-electron wave
packet, we establish the apparent ``time zero'' when each of the two
photoelectrons leaves the atom. This time depends sensitively on the
photon energy and the energy sharing between the photoelectrons.  

To facilitate an individual attosecond streaking in a two-electron
ionization process, it was suggested in 
Ref~\cite{PhysRevLett.94.213001} to direct emitted electrons parallel
and perpendicular to the NIR field to respectively maximize and
minimize its streaking effect. We adopt this configuration and direct
two photoelectrons perpendicular to each other $\k_1\perp\k_2$. We
also distinguish the reference photoelectron, the one which would be
streaked, and its spectator counterpart, which influences the
reference photoelectron via their mutual Coulomb interaction.
%
%
For simplicity of our analysis, both photoelectrons
are kept in the same $xz$ plane with the polarization vector of the
XUV radiation  directed along the $\z$ axis.

The time-dependent calculation of DPI of He was performed by radial
grid integration of the TDSE using the Arnoldi-Lanczos method
\cite{park:5870}.
We used the linearly polarized XUV pulse $\E(t)=E_0 \,g(t)
\cos{\omega t}$ with the envelope function  $g(t)=\cos^2(\pi t\slash
2T_1)$ centered at $t=0$, which we take as the physical ``time zero''.
The peak field strength was $E_0=0.1$~a.u., carrier frequency
$\omega=106$~eV and $T=2\pi\slash\omega=39$~as.  The pulse was turned off
outside the interval $\pm T_1$, where $T_1=4T$.
%


%
The field-free solution of the TDSE for $t>T_1$ was used to extract
information about the DPI process with asymptotic photoelectron
momenta $\k_1$, $\k_2$.  This task was achieved by tracing time
evolution of the wave packet state $\Psi_1(\r_1,\r_2,t)=\hat
P_{\k_1,\k_2 } \Psi(\r'_1,\r'_2,t)$, where the kernel of the
projection operator was constructed as
\ba
\langle \r'_1,\r'_2|\hat P_{\k_1,\k_2}|\r_1,\r_2\rangle&=&\int_{\Omega}
\Psi^-_{\q_1}(\r_1)
\Psi^-_{\q_2}(\r_2)\\
&&\hs{-5mm}\times
\Psi^-_{\q_1}(\r'_1)^*
\Psi^-_{\q_2}(\r'_2)^*
d{\q}_1 d{\q}_2
\nn
\ .
\label{proj}
\ea
Here $\Psi^-_{\k_i}(\r_i)$ are one-electron scattering states with
the ingoing boundary condition describing a photoelectron moving in the
Coulomb field with $Z=2$.  The integration region is defined as
$\Omega=\Omega_1 \bigotimes \Omega_2$, where $\Omega_1$ and $\Omega_2$
are spheres in momentum space centered around the momentum vectors
$\k_1,\k_2$ so that $|\bm q_i-\k_i|< 0.25k_i$.

The wavepacket state $\Psi_1(t)$ can be expanded over the set of the
double continua  states of the He atom as
\be
\Psi_1(\r_1,\r_2,t) \!\!=\!\!
\int \!\!d{\q_1} d{\q_2}
f({\q_1},{\q_2}) \Psi_{\q_1,\q_2}^-(\r_1,\r_2) e^{-i Et}
\ ,
\label{asf}
\ee
%
%
%
where $E=q_1^2\slash2 +q_2^2\slash2$.
When both $r_1,r_2$ and $r_{12}$ are large, 
\be
\Psi_{\q_1,\q_2}^-(\r_1,\r_2) \propto 
\exp[i(\q_1\cdot\r_1+\q_2\cdot\r_2+\gamma)]
\ ,
\ee
where $\gamma$ is the Redmond logarithmic  phase  \cite{0022-3700-15-20-006}.
This leads to the following asymptotic expression
\ba
\Psi_1(\r_{1,2}\to\infty,t>T_1) &\asymp& 
\int_{\Omega} d{\q_1} d{\q_2} 
|f({\q_1},{\q_2})| 
\\&&\hs{-35mm}\times
\exp\Big\{i[\arg f(\q_1,\q_2)+
{\q_1}\r_1+{\q_2}\r_2+\gamma
-E\,t
]\Big\}
\nn
\label{asf1}
\ea
The center of the wave packet
moves in such a way that its phase is stationary with respect to 
both ${\q_1}$ and ${\q_2}$ at the points 
$\k_1$,$\k_2$ of the center of the wavepacket:
\be
(d\slash d\k_i)
\left[\arg f(\k_1,\k_2)+\k_1\r_1+\k_2\r_2 +\gamma
-E\,t
\right]
=0
\ ,
\label{sp}
\ee
where $\k_i$ is either $\k_1$ or $\k_2$. This gives asymptotic
equations for the electron trajectories:  
\ba
\label{straight}
\r_i \asymp \k_i 
\left[
t -{d\gamma\slash dE_i}-{d\arg f(\k_1,\k_2)\slash dE_i}
\right]
\ .
\label{et}
\ea
The term containing  derivative of $\gamma$ gives logarithmic (with
$t$) corrections to the electron trajectory, which can thus be
represented asymptotically as
\be
r_i(t)-k_i t-r_{1i}(t) \asymp k_i t_{0i}
\ . 
\label{et1}
\ee
Here $t_{0i}= d\arg f(\k_1,\k_2)\slash dE_i$ are the time delays and
$r_{1i}(t)$ are known functions which vary logarithmically with $t$.

\begin{figure}[h]
\hs{-1.25cm}
\epsfxsize=8cm
\epsffile{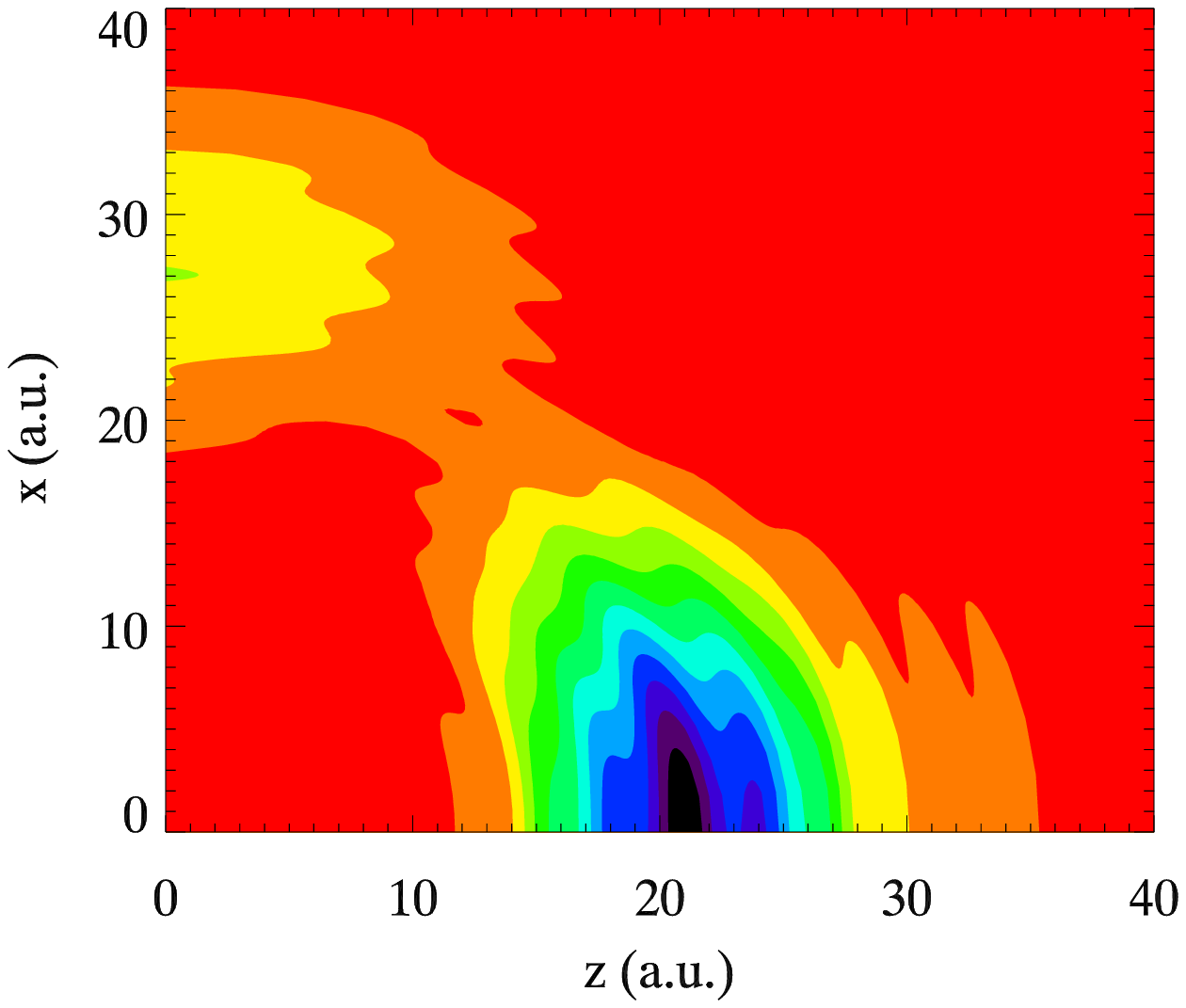}
\vs{-1cm}

\epsfxsize=6cm
\epsffile{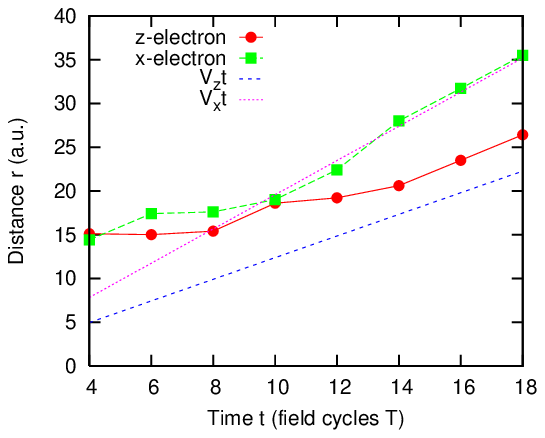}
%

\epsfxsize=6cm
\epsffile{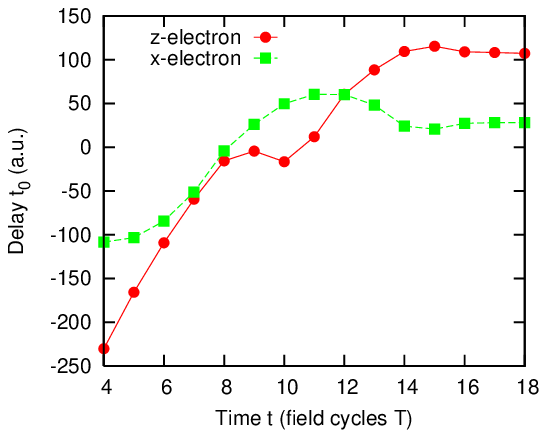}

\caption{
Time evolution of the two-electron wave packet for 8~eV and 20~eV
photoelectrons propagating along the $z$ and $x$ axes,
respectively. Top: the electron density plot in the $xz$ plane at
$t=14T$. Middle: trajectories of both photoelectrons as functions of
time measured in numbers of field periods. The straight lines
visualize the free propagating $V_zt$ and $V_xt$. Bottom: the effective
time delay computed from the LHS of \Eref{et1}.  }
\vs{-0.5cm}
\label{Fig1}
\end{figure}

\begin{figure}
\vs{3cm}
\epsfxsize=5cm
\epsffile{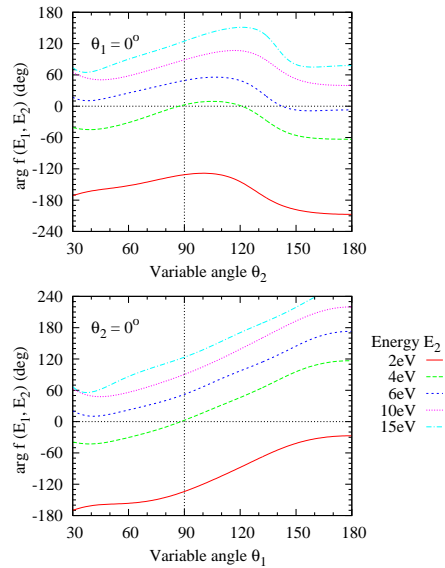}

\caption{
The phases of the DPI amplitude $f(E_1,E_2)$ are plotted for
$E_2=2$, 4, 6, 10, 15~eV and $E_1=40$~eV. On the top panel, the
fast photoelectron angle $\theta_1=0^\circ$ is fixed and the slow
electron angle $\theta_2$ is variable. On the bottom panel it is vise
versa.
}
\vs{-0.5cm}

\label{Fig2}
\end{figure}

We apply these asymptotic formulas to describe time evolution of the
maxima of the electron density defined as $\rho(\r,t)=\int
|\Psi_1(\r_{1,2},t)|^2 (\delta(\r-\r_1)+\delta(\r-\r_2))\ d\r_1d\r_2$
and $\Psi_1(\r_{1,2},t)$ provided by the TDSE solution.  
As an illustration of our technique, we consider a DPI process in
which one photoelectron escapes with energy 8~eV along the $z$-axis
and the other with energy 20~eV along the $x$-axis.
The snapshot of the electron density in the $xz$ plane corresponding
to the moment of time $t=14 T$ (10 field cycles elapsed after the end
of the XUV pulse) is shown on the top panel of \Fref{Fig1}.
The figure exhibits clearly the two well formed maxima corresponding
to the center of the wavepacket $\Psi_1(t)$ propagating in the $x$ and
$z$ directions.  A sequence of such snapshots is taken with an
interval of $2T$ and the maxima of the electron density are traced in
time. This procedure defines the trajectories $r_i(t)$ for both
photoelectrons which are exhibited in the middle panel of \Fref{Fig2}.
The free propagation is visualized to the straight lines $V_zt$ and
$V_xt$.
Knowing the trajectories, we can compute the left-hand side of
\Eref{et1} which is plotted on the bottom panel of the
\Fref{Fig1}. At sufficiently large time, this quantity should
 approach the constant values of the time delays $t_{0i}$ for both
 photoelectrons. These values are 107~as and 28~as for the 8~eV and
20~eV photoelectrons, respectively. We ran an analogous simulation
with the directions of the fast and slow photoelectrons being swapped
and obtained very similar time delay figures. We also explored the
case of the equal energy sharing of both photoelectrons and obtained
the time delay of about 55~as for the 14~eV photoelectrons.

To relate the time delay to the phases of the DPI amplitudes, we
employ the lowest order, with respect to the field, perturbation
theory (LOPT) on the basis of channel states of the convergent
close-coupling (CCC) method \cite{PhysRevA.54.R995}. This basis is
composed of the products of a Coulomb wave $\chi^-_\k$ (for the faster
of the two photoelectrons) and a positive energy pseudostate $ \phi_i$
(for the slower photoelectron) which is obtained by diagonalizing the
He$^+$ Hamiltonian in a truncated Laguerre basis. Thus, we write
\ba
\Psi(\r_1,\r_2,t) &=& \!\!
-i
\sum_{i} 
\isum_{\k} d^3k \ 
\la \chi^-_\k  \phi_i\, | D |\Phi_0\ra \
\chi^-_\k (\r_1)\ \phi_i(\r_2)
\nn\\ &&\hs{15mm}\times
e^{-iE_{ik} t}
\tilde  \E(E_{ik}-E_0)
\ .
\ea
Here 
$
\tilde \E(\omega)=
\int_{-\infty}^{\infty} e^{i\omega\tau} \E(\tau)\ d\tau
$
is the Fourier transform of the XUV field, $D$ is the two-electron
dipole operator, $E_{ik}=k^2\slash2+\epsilon_i$ and
$k^2\slash2\ge\epsilon_i>0$.  By projecting the positive energy
pseudostate onto the matching energy Coulomb wave
$k_2^2\slash2=\epsilon_{n_2l_2}$, we restore the continuum
normalization and phase and write a two-electron wave packet in the
form of a partial wave expansion:
\ba
\Psi_1(\r_1,\r_2,t) &=& 
i
\sum_{l_1l_2} 
\isum_{k_1} dk_1 \ 
\tilde D_{l_1l_2}(k_1,k_2) 
\\ &&\hs{-2cm}
\times
 R_{k_1l_1}(r_1) R_{k_2l_2}(r_2)
\Y^{\,l_1l_2}_1(\hat\r_1,\hat\r_2)
e^{-iE t}
\tilde \E(E-E_0) 
\nn
\ea
Here $\Y^{\,l_1l_2}_1$ is a bipolar harmonic and $E=k_1^2\slash2
+k_2^2\slash2$. The two-electron dipole matrix element is defined as
\ba
\label{DME}
\tilde D_{l_1l_2}(k_1,k_2) &=& (-i)^{l_1+l_2} \
e^{i[\delta_{l_1}(Z=1)+\delta_{l_2}(Z=2)]}
\nn\\&&\hs{3mm}\times \ D_{l_1l_2}(k_1n_2)
\langle l_2k_2\parallel l_2n_2\rangle
\ea
with $D_{l_1l_2}(k_1n_2)$ to be found by integrating the bare dipole
matrix element with the half on-shell $T$-matrix
\cite{PhysRevA.54.R995}.


Knowing the asymptotics of the radial orbitals
$
R_{kl} \propto
\sin{\big[kr+\delta_l(k)+{1\slash k}\ln(2kr)-l\pi\slash 2}
\big]
$
and applying the usual saddle-point approximation, we arrive to
\Eref{sp} with the following definition of the DPI amplitude:
\be
f(\k_1,\k_2) = 
\sum_{l_1l_2} 
\tilde D_{l_1l_2}(k_1,k_2) 
{\cal Y}^{\,l_1l_2}_1(\hat\k_1,\hat\k_2)
\vs{-3mm}
\ee
To obtain the time delay $t_{0i}= d\arg f(\k_1,\k_2)\slash dE_i$, we
have to take the derivative over the energy of the corresponding
photoelectron. However, these derivatives are only defined for
$E_1> E\slash2$ and $E_2\le E\slash2$. So neither of photoelectrons can
serve as the reference electron in the full excess energy range. In
the following, we choose the photoelectron 2 to be the reference
one. We compute the time delay $t_{02}$ which is defined for $E_2\le
E\slash2$, and continue it analytically past the mid excess energy
point $E_1> E\slash2$.

An example of the phase plot of the DPI amplitude is given in
\Fref{Fig2}. In the $xz$ plane, geometry of the two-electron escape
is fully defined by the two azimuthal angles $\theta_1,
\theta_2$.  On the top panel of \Fref{Fig2}, the phase $\arg
f(E_1,E_2)$ is plotted for the slow electron energies $E_2=2$, 4, 6,
10 and 15~eV and variable angle $\theta_2$ whereas the fast electron
energy $E_1=40$~eV and its angle $\theta_1=0^\circ$ are fixed. On the
bottom panel, directions of the photoelectrons are swapped and the
angle of the slow photoelectron is fixed at $\theta_2=0^\circ$ while the
angle of the fast photoelectron $\theta_1$ varies. The phases of the
DPI amplitudes displayed in \Fref{Fig2} can be used to obtain the
timing information of the two complementary processes in which the
slow reference photoelectron is directed along with (bottom panel) and
perpendicular to (top panel) the XUV field.

From inspection of \Fref{Fig2}, we see that the phases of the DPI
amplitudes depend sensitively on the mutual photoelectron orientation.
However, the spacing between the various $E_2$ phase curves does not
change significantly with the variable photoelectron angle $\theta_2$.
So the energy derivative of the phase of the DPI amplitude, and hence
the effective time delay, does not change very much with the relative
orientation of the photoelectrons. 
%

\begin{figure}[h]
\vs{4cm}
\epsfxsize=5cm
\epsffile{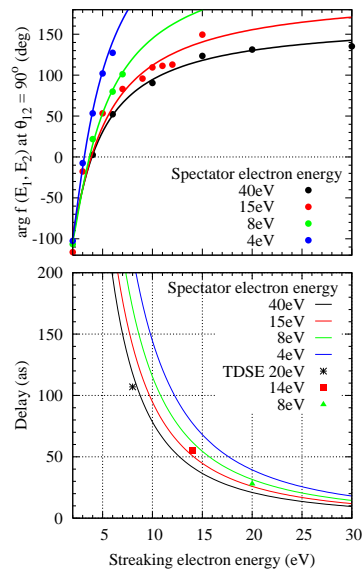}

\caption{
Top:  cumulative phase plot $\arg
f(E_1,E_2)$ for various combinations of the streaked and spectator
electron energies taken at the perpendicular orientation
$\theta_{12}=90^\circ$. 
Bottom: The time delay $t_{02}= d\arg f(\k_1,\k_2)\slash dE_2$ for
various parallel and perpendicular electron energies.  The values of
time delay obtained from the solution of TDSE are marked by the points
colored respectively to match the combination of $E_1$ and $E_2$ from
the CCC calculation.
\vs{-5mm}
}
\label{Fig3}
\end{figure}

The vertical line on the both panels of \Fref{Fig2} marks the mutual
angle of the photoelectrons $\theta_{12}=90^\circ$. The phases on both
panels are very similar for this orientation which means that the time
delay of the reference photoelectron does not depend significantly on
its orientation relative to the XUV field. We have already
acknowledged this fact when analyzing the TDSE time delay results.

Weak sensitivity of the time delay to the field orientation can be
understood from the general parametrization of the DPI amplitude:
\ba
f(\k_1,\k_2) &=&  [\cos\theta_1+\cos\theta_2]  \, \M^g(E_1,E_2,x)
\\&&\hs{0.5cm}  +
[\cos\theta_1-\cos\theta_2]  \, \M^u(E_1,E_2,x)
\ .
\nn\ea
Here $x=\cos\theta_{12}=\cos(\theta_2-\theta_1)$  and the
complex gerade $\M^g$ and ungerade $\M^u$ amplitudes possess the
exchange symmetry
$\ds
\M^{g\slash u}(E_1,E_2) = \pm \M^{g\slash u}(E_2,E_1)
 .
$
Even for the most severe energy sharing $E_2\ll E_1$,
the gerade amplitude is still strongly dominant
$
|\M^g|\gg |\M^u|
$
\cite{PhysRevA.65.022708}.
Therefore, unless the photoelectrons are anti-parallel and the
kinematic factor accompanying the gerade amplitude tends to zero, its
contribution is dominant and it is $\M^g$ that determine the overall
DPI phase. This means that the timing measurement at perpendicular
photoelectron orientation can only deliver the $\M^g$ phase.  An
analogous measurement for $\M^u$ would require the anti-parallel
orientation which is not practicable for individual photoelectron
streaking. However, since the relative phase of $\M^{g\slash u}$ can
be determined independently,  knowing the  $\M^g$ phase 
 will immediately deliver the missing  phase of $\M^u$.

On the top panel of \Fref{Fig3}, we show a cumulative phase plot $\arg
f(E_1,E_2)$ for various combinations of the reference  and
spectator  electron energies taken at the perpendicular
orientation $\theta_{12}=90^\circ$. 
The raw CCC data, marked by the points, are only available for the
$E_2\le E_1$. To obtain the phases across the whole excess energy
range, we fit the raw CCC data with a rational function and continue
it analytically past the mid excess energy point. The energy
derivative of this function, calibrated in units of time delay, is
presented on the bottom panel of
\Fref{Fig3}. The values of time delay obtained from the solution of
TDSE are marked by the points colored accordingly to match the
combination of $E_1$ and $E_2$ from the CCC calculation. We observe
that for these particular combinations of the reference and spectator
photoelectron energies, the TDSE and CCC time delays are quite close.

The time delay of the reference photoelectron varies very rapidly with
its energy but depends much weaker on the energy of the spectator
electron. When the energies of the both electrons are low, the time
delay is particularly large reaching few hundred of attoseconds. In
the opposite limit of large reference electron energy, the time delay
becomes small. More importantly, it does not vary significantly with
the spectator electron energy. Physically, this regime corresponds to
the shake-off  mechanism of DPI in which the fast photoelectron
absorbs the whole of the photon energy and angular momentum and the
slow photoelectron is subsequently shaken off into the continuum
\cite{PhysRevLett.89.033004}.  The fast photoelectron leaves the atom
without any significant delay, but emission of the slow photoelectron
is delayed considerably. This delay becomes particularly large when
the energy of the both photoelectrons is small and they are able to
interact for a long time. In this regime, the main mechanism of DPI is
the knock-out process in which the primary photoelectron impinges on
the ion and knocks out the secondary electron into the continuum.

In conclusion, we perform the timing analysis of the two-electron
emission from the He atom which is subjected to a very short XUV
pulse. We employ an explicit time-dependent treatment of the DPI
process by seeking solution of the TDSE. We complement this procedure
by the LOPT treatment which allows us to connect the time delay with
the energy dependent phase of the DPI amplitude, the latter being
evaluated within the CCC method. This opens up a possibility of a
complete DPI experiment in which both the magnitudes and phases of the
symmetrized DPI amplitudes can be determined. Such an experiment will
require an attosecond streaking measurement on one of the two
photoelectrons which can be performed in a close to perpendicular
orientation of the photoelectrons. To our best knowledge, except for
a very recent report \cite{1367-2630-12-10-103024}, this is the first
practical attosecond streaking measuring scheme suggested for a DPI
process.

The authors acknowledge support of the Australian Research Council in
the form of the Discovery grant DP0771312.  Resources of the National
Computational Infrastructure (NCI) Facility were employed.
\vs{-5mm}


\vs{-5mm}
\end{document}